\documentclass[letterpaper]{article} 
\usepackage{aaai2026}  
\usepackage{times}  
\usepackage{helvet}  
\usepackage{courier}  
\usepackage[hyphens]{url}  
\usepackage{graphicx} 
\usepackage{natbib}  
\usepackage{caption} 
\usepackage{algorithm}
\usepackage{algorithmic}

\usepackage{times}
\usepackage{soul}
\usepackage{url}
\usepackage[utf8]{inputenc}
\usepackage{amsmath}
\usepackage{amsthm}
\usepackage{booktabs}
\usepackage{algorithm}
\usepackage{algorithmic}
\usepackage[switch]{lineno}
\usepackage{xcolor}
\usepackage{amssymb}
\usepackage{subcaption}
\usepackage{booktabs}
\usepackage{siunitx}
\usepackage{makecell}
\usepackage{graphicx} 

\usepackage{newfloat}
\usepackage{listings}
\DeclareCaptionStyle{ruled}{labelfont=normalfont,labelsep=colon,strut=off} 
\floatstyle{ruled}
\newfloat{listing}{tb}{lst}{}
\floatname{listing}{Listing}
\theoremstyle{definition}
\newtheorem{definition}{Definition}[section]

 \nocopyright 

\title{Fairness Aware Reinforcement Learning via Proximal Policy Optimization}
\author{
    Gabriele La Malfa\textsuperscript{\rm 1}, Jie M. Zhang\textsuperscript{\rm 1}, Michael Luck\textsuperscript{\rm 2}, Elizabeth Black\textsuperscript{\rm 1}
}
\affiliations{
    \textsuperscript{\rm 1}King’s College London\\
    \textsuperscript{\rm 2}University of Sussex\\
    gabriele.la malfa@kcl.ac.uk, jie.zhang@kcl.ac.uk, michael.luck@sussex.ac.uk, elizabeth.black@kcl.ac.uk

}

\usepackage{bibentry}

\begin{document}

\maketitle

\begin{abstract}
Fairness in multi-agent systems (MAS) focuses on equitable reward distribution among agents in scenarios involving sensitive attributes such as race, gender, or socioeconomic status. This paper introduces fairness in Proximal Policy Optimization (PPO) with a penalty term derived from a fairness definition such as demographic parity, counterfactual fairness, or conditional statistical parity. 
The proposed method, which we call Fair-PPO, balances reward maximisation with fairness by integrating two penalty components: a retrospective component that minimises disparities in past outcomes and a prospective component that ensures fairness in future decision-making.
We evaluate our approach in two games: the Allelopathic Harvest, a cooperative and competitive MAS focused on resource collection, where some agents possess a sensitive attribute, and HospitalSim, a hospital simulation, in which agents coordinate the operations of hospital patients with different mobility and priority needs. Experiments show that Fair-PPO achieves fairer policies than PPO across the fairness metrics and, through the retrospective and prospective penalty components, reveals a wide spectrum of strategies to improve fairness; at the same time, its performance pairs with that of state-of-the-art fair reinforcement-learning algorithms. Fairness comes at the cost of reduced efficiency, but does not compromise equality among the overall population (Gini index). These findings underscore the potential of Fair-PPO to address fairness challenges in MAS.\footnote{Code available here: \\https://github.com/gabrielelamalfakcl/fairness-aware-ppo}
\end{abstract}

\section{Introduction}~\label{Introduction}  
In multi-agent systems (MAS), agents interact in an environment to pursue individual or shared goals. Fairness in MAS focuses on whether the reward distribution mechanisms, driven by agent decisions or other processes, treat agents fairly. For instance, fair reinforcement learning explores methods to promote fairness by enabling agents to learn a fair policy~\cite{Reuel2024}; fair division addresses fair resource allocation~\cite{Lindner2016,Amantidis2023}; and negotiation designs methods for fair bargaining resolution~\cite{Guth2014,Debove2016}.

On the other hand, in human society, fairness is framed in terms of inequality or discrimination between privileged and disadvantaged groups. Sensitive attributes, such as race, gender and socioeconomic status, define subgroups historically marginalised in workplaces, healthcare, education, and politics.\footnote{Throughout the paper, we use the term `sensitive attribute' instead of `protected characteristic' to avoid any confusion with the legal meaning reported, for example, in the UK Equality Act.} 
To enhance fairness, individuals (are often nudged to) adjust their behaviour towards those holding sensitive attributes. For example, giving up a seat on public transport for an elderly person illustrates a behavioural adjustment to promote fairness.
For this reason, integrating fairness into agents' policies has been an area of growing investigation~\cite{Reuel2024}. 

Foundational works in social sciences~\cite{Griesinger1973,Liebrand1984} have identified agents' attributes as a crucial factor influencing fairness outcomes in games. In this sense, inspired by algorithmic fairness~\cite{Mitchell2021,Castelnovo2022}, we propose sensitive attributes as characteristics that should not affect an agent's expected reward. We apply metrics from the algorithmic fairness literature, specifically demographic parity, counterfactual fairness, and conditional statistical parity, to the MAS context and use them to constrain agent behaviour and obtain fair policies. Building on gradient-based algorithms in reinforcement learning, and inspired by the work of Zhang et al.~\cite{Zhang2022}, we propose a fairness-aware Proximal Policy Optimization (PPO)~\cite{Schulman2017A} method, which we call Fair-PPO, that improves policy fairness. We modify the PPO objective function to include a penalty term derived from a fairness metric, allowing multi-objective optimisation of the policy that accounts for both performance and fairness. 


Our proposed penalty has two components. The first component penalises total reward disparities between agents that differ by a sensitive attribute by looking at past outcomes.
The second component penalises disparities in the expected rewards as per the estimate of the value function of each agent. In other words, the first component is retrospective, addressing disparities in past outcomes, while the second is prospective, encouraging fairness in the agent's future expectations and decision-making.

In summary, the main contribution of this work is the novel Fair-PPO reinforcement learning algorithm, which extends PPO with the retrospective and prospective penalty components.
We perform experiments in a version of the Allelopathic Harvest (AH) game~\cite{Leibo2019}, a MAS that combines cooperation and competition in resource collection, where two groups of agents with different preferences regarding available resources navigate a dynamic environment. Agents are distinguished according to whether they hold some sensitive attribute: agents with this attribute move more slowly and so are potentially disadvantaged in resource collection. 
Our evaluation of Fair-PPO extends to HospitalSim (HS), a novel hospital simulation. This MAS integrates classification, matching, and resource allocation as agents manage patients with different mobility and priority needs moving through various hospital areas.
 
Our experiments show that: 
(i) Fair-PPO produces fairer policies than PPO across the fairness metrics we propose and, through the retrospective and prospective penalty components, reveals a wide spectrum of strategies to improve fairness; (ii) fairness comes at the cost of efficiency but does not compromise equality among the overall population of agents, as measured by the Gini index; and (iii) Fair-PPO is well-suited for finding fair policies in a range of simulation types, from cooperative and competitive to coordination-based scenarios.

The paper is structured as follows: in Section~\ref{RelatedWork}, we review the literature on fairness in MAS, fairness in reinforcement learning and algorithmic fairness. Section~\ref{Preliminaries} introduces preliminary concepts of MAS and PPO. In Section~\ref{Fair-PPO}, we detail the fairness metrics and the penalty component integration into PPO. Sections~\ref{Experiments} and~\ref{Results} focus on evaluating our approach, presenting experimental results using AH and HS.

\section{Related Work}~\label{RelatedWork}
Our work is inspired by algorithmic fairness in seeking to contribute to fair reinforcement learning literature in MAS. We review recent prominent works in these fields.


\subsection{Fairness Measures in MAS}
In MAS, fairness has been evaluated through methods tailored to the system design and objectives.
A widely accepted and intuitive notion is \textit{proportionality}: fairness is determined by the \textit{proportion} of reward allocated to an agent or group compared to others. Proportionality is applied in many MAS where a resource is allocated to multiple agents, such as the ultimatum game~\cite{Guth2014,Debove2016} or fair division games~\cite{Lindner2016,Amantidis2023,Murhekar2024}.   
Beyond proportionality, \textit{envy-freeness} ensures that no agent prefers another's allocation, and \textit{maximin share fairness} guarantees that each agent receives a share at least as good as they could secure by dividing resources themselves~\cite{Lipton2004,Budish2011,Caragiannis2019}. 
Alternative fairness measures propose to allocate rewards based on an agent's merit~\cite{Joseph2016}, enhancing \textit{treatment equality} by equalising error rates across groups~\cite{Liu2017}, or minimising \textit{regret}, which captures the cost of deviating from the optimal trade-off between fairness and efficiency~\cite{Li2020,Patil2021,Jones2023,Barman2023}. 
Last but not least are wealth distribution metrics such as \textit{social welfare}~\cite{Kaplow2003,Brânzei2017,Høgsgaard2023}, which measure the total utility of all individuals based on a chosen outcome, and measures of inequality such as the Gini index~\cite {Farris2010}.

Our approach is based on the principle of ensuring a fair distribution of rewards among agents, similar to proportionality and equitable wealth distribution. However, the fundamental distinction lies in the basis of group formation through the sensitive attributes that permit the use of specialised fairness metrics from the algorithmic fairness literature to confront precise biases. 


\subsection{Algorithmic Fairness}
Algorithmic fairness addresses bias and discrimination in decision-making systems across domains such as justice~\cite{Berk2019}, education~\cite{Baker2021}, credit scoring~\cite{Kozodoi2022}, and healthcare~\cite{Vyas2020}, \cite{Giovanola2022}, with a focus on protected attributes characterising discriminated groups.
Fairness metrics are classified into group and individual categories. Group fairness metrics include \textit{demographic parity}~\cite{Kamishima2012} and \textit{equalised odds}~\cite{Hardt2016}, which use confusion matrix rates, while \textit{calibration-based metrics} evaluate prediction accuracy relative to group membership~\cite{Chouldechova2016}. Individual fairness, such as \textit{counterfactual fairness}~\cite{Kusner2018}, assesses consistency across factual and counterfactual scenarios.

\subsection{Fairness and Reinforcement Learning}
Reinforcement learning traditionally focuses on learning policies that maximise expected rewards~\cite{Sutton2018}. However, this raises fairness concerns, as it can perpetuate biases and violate fairness and legal principles~\cite{Jabbari2017}. To address these issues, fairness constraints can be added to the optimisation process. For example, Siddique et al.~\cite{Siddique2020} and Zimmer et al.~\cite{Zimmer2021} define fairness as finding solutions that are \textit{efficient} (benefiting everyone without waste), \textit{impartial} (treating identical agents equally), and \textit{equitable} (helping those who are worse off).

Chen et al.~\shortcite{Chen2021} propose adjusting rewards through a multiplicative weight to achieve $\alpha$\textit{-fairness}, while Zhang et al.~\shortcite{Zhang2014} implement \textit{maximin fairness} to optimise the worst-performing agent's outcome. Other works explore fairness across agent groups, including \textit{demographic parity}~\cite{Jiang2019,Wen2021,Chi2022}. Some contributions address real-world complexities, such as agents with differing characteristics or preferences, necessitating tailored fairness mechanisms~\cite{Yu2023,Ju2024}. 

To contextualise Fair-PPO as an algorithm to enhance fairness in decentralised multi-agent reinforcement learning, we benchmark it to FEN~\cite{Jiang2019} and SOTO~\cite{Zimmer2021}. While Fair-PPO, FEN and SOTO all provide solutions to balance efficiency and fairness, they diverge significantly in their architectural and optimisation strategies.
Fair-PPO incorporates a penalty term based on tailored fairness definitions, such as demographic parity, into its optimisation objective. 
FEN adopts a hierarchical learning system and defines a fair-efficient reward to guide agents toward fair outcomes. SOTO, instead, optimises a social welfare function, and it is flexible to fairness metrics adaptation.


\section{Preliminaries} ~\label{Preliminaries}
In this section, we first define the elements composing a MAS and then define gradient-based policies and PPO.

\subsection{Multi-Agent Systems}
A MAS consists of multiple agents acting in an environment to achieve their goals. Let $\mathcal{N} = \{1, \ldots, n\}$ be the set of $n$ agents.
We denote a MAS as $\mathcal{M} = (S, s_0, A, \mathcal{N}, At, P)$, where $S$ is the set of possible environment states, $s_0 \in S$ is the initial state, $A = A_1 \times \ldots A_n$ is the joint action space, where $A_i$ is the set of actions available to agent $i \in \mathcal{N}$; $At$ is the set of attributes available to the agents and we designate a single binary sensitive attribute $z \in At$ for fairness considerations; $P: S \times A \rightarrow \Delta(S)$ is the non-deterministic state transition function, which maps a state-action pair to a probability distribution over the next states.

We define an agent $i \in \mathcal{N}$ as a tuple of attributes, policy and individual reward function. For the designated sensitive attribute $z \in At$, we denote its value for agent $i$ as $z_i \in \{0, 1\}$. This allows us to partition the agents into a sensitive group $\mathcal{N}_1=\{i \in \mathcal{N} \mid z_i=1\}$ and a non-sensitive group $\mathcal{N}_0=\{i \in \mathcal{N}\mid z_i=0\}$. 
The stochastic policy $\pi_i: S \rightarrow \Delta(A_i)$ of agent $i$ is a function mapping any given state $s \in S$ to a probability distribution over the agent's actions $A_i$. We denote the probability of taking an action $a_i$ in state $s$ as $\pi_i(a_i \mid s)$ and the joint policy as $\pi = \{\pi_1, \ldots, \pi_n\}$. The reward function for agent $i$ from state $s_t$ to $s_{t+1}$, due to the joint action $a_t$, is $R_i: S \times A \times S \rightarrow \mathbb{R}$. In this way, an agent $i$ receives a reward $r_{t+1, i} = R_i(s_t, a_t, s_{t+1})$.

We define a trajectory or episode as a sequence of states, actions and rewards $\tau = (s_0, a_0, r_1, s_1, a_1, r_2, \ldots)$, where $a_t = (a_{t,1}, \ldots a_{t, n})$ is the joint action at time $t$ and $r_{t+1}= (r_{t+1,1}, \ldots, r_{t+1,n})$ is the resulting joint rewards. The total reward achieved by an agent $i$ over a trajectory $\tau$ of length $T$ is $G_i(\tau) = \sum_{t=1}^T r_{t,i}$.

The probability of a trajectory $\tau = (s_0, a_0, \ldots, s_T)$ occurring under a joint policy $\pi$ is $p(\tau \mid \pi) = p(s_0) \prod \limits_{t=0}^{T-1} P(s_{t+1} \mid s_t, a_t) \prod \limits_{i=1}^{n} \pi_i(a_{t, i}\mid s_t)$, which is the product of the initial state probability, the environment transition probability and the policy probabilities at each step.
The expected reward of an agent $i$ is $J_i(\pi) = \sum_{\tau} p(\tau \mid \pi) G_i(\tau)$, which is the sum of the total rewards $G_i(\tau)$ from every possible trajectory multiplied by the probability of that trajectory.

\subsection{Proximal Policy Optimization}~\label{PPO}
In reinforcement learning, policy-gradient methods optimise the parameters $\theta$ of a policy $\pi_{\theta}$ to maximise the agent's objective function. These parameters are updated through gradient ascent. 
To improve the stability of the learning process, modern algorithms implement methods that constrain the policy update, like Trust Region Policy Optimization~\cite{Schulman2017B}.
PPO works with Clipped Surrogate Objective (CLIP), limiting, within a small range, the change in the probability ratio of actions between the old and new policies. 

PPO integrates policy optimisation and value function accuracy into a single objective function as follows:
\begin{equation} \label{LossFunc}
    L_t(\theta_i) = \hat{\mathbb{E}}_t \left[L_t^{\text{CLIP}}(\theta_i) - c_1 L_t^{\text{VF}}(\theta_i) + c_2 H[\pi_{\theta_i}](s_t)\right]
\end{equation}
where $t$ is a timestep, $i$ is the agent and $c_1$ and $c_2$ are weighting coefficient.
The objective is composed of the following three components.

The Clipped Surrogate Objective $L^{\text{CLIP}}$ constrains the policy update to improve the stability: 
\begin{align*}
L_t^{\text{CLIP}}(\theta_i) =  \hspace{-1cm} & \\
& \begin{aligned}
\min\left(\psi_t(\theta_i)\hat{A}_{t,i}, \ \text{clip}\left(\psi_t(\theta_i), 1 - \epsilon, 1 + \epsilon\right)\hat{A}_{t,i}\right)
\end{aligned}
\end{align*}
where $\psi_t(\theta_i)$ is the ratio of the probability of taking action $a_t$ under the new policy to the probability of taking it under the old policy; $\hat{A}_{(t, i)}$ is the advantage function for agent $i$ at step $t$, which estimates how much better the action $a_t$ is compared to the average value of the state.

The Value Function Loss improves the accuracy of the policy's value estimation:
\begin{equation} \nonumber
       L_t^{\text{VF}}(\theta_i) = \left(V_{\theta_i}(s_t) - G_{t,i}\right)^2
\end{equation}
where $V_{\theta_i}(s_t)$ is the estimate of the return for state $s_t$, and $G_{t, i}$ is the empirical discounted rewards for agent $i$ from state $s_t$ onwards.

Finally, the Entropy Bonus encourages exploration by promoting more diverse action selection:
\begin{equation*}
H[\pi_{\theta_i}](s_t) = 
-\sum_{a_i \in A_i} \pi_{\theta_i}(a_i \mid s_t) \log \pi_{\theta_i}(a_i \mid s_t).
\end{equation*}

\section{Fair-PPO}~\label{Fair-PPO}
In this section, we formalise demographic parity, counterfactual fairness and conditional statistical parity in MAS when sensitive attributes are involved; then, we detail the demographic parity-based penalty and integrate it as a constraint into the PPO objective function. The objective function formulations incorporating counterfactual fairness and conditional statistical parity penalties can be found in Section A of the supplementary material.

\subsection{Fairness Metrics in MAS}~\label{FairnessMetrics}
Fair-PPO incorporates the notions of demographic parity, counterfactual fairness and conditional statistical parity as a penalty term into the gradient optimisation. These definitions compare the average expected returns of agent groups partitioned by the sensitive attribute $z$.

Let $\pi$ be the joint policy. We define the average expected rewards for any subgroup of agents $\mathcal{G} \subseteq \mathcal{N}$ as:
$\bar{J}_{\mathcal{G}}(\pi) = \frac{1}{|\mathcal{G}|} \sum_{i \in \mathcal{G}} J_i(\pi)$.

\begin{definition}[\textbf{Demographic Parity}] \label{Demographic Parity}
In a MAS $\mathcal{M}$, demographic parity is satisfied if the average expected rewards are the same for both sensitive and non-sensitive groups $\mathcal{N}_1$ and $\mathcal{N}_0$, i.e., 
$\bar{J}_{\mathcal{N}_1}(\pi) = \bar{J}_{\mathcal{N}_0}(\pi)$.
When demographic parity does not hold, we quantify the disparity as the absolute difference between these averages:
\begin{equation} \label{EqDP}
\Delta DP(\pi) = \left| \bar{J}_{\mathcal{N}_1}(\pi) - \bar{J}_{\mathcal{N}_0}(\pi) \right|
\end{equation}
\end{definition}




\begin{definition}[\textbf{Counterfactual Fairness}] \label{Counterfactual Fairness}
Consider a factual system $\mathcal{M}$ with joint policy $\pi$, and a counterfactual system $\mathcal{M}^\prime$ which is identical to $\mathcal{M}$ except that the sensitive attribute $z$ is flipped for all agents. Let $\pi^\prime$ be the joint policy in $\mathcal{M}^\prime$. Counterfactual fairness is satisfied if the expected return for every agent remains unchanged, i.e., 
$J_i(\pi)=J_i(\pi^\prime)$ $\forall i \in \mathcal N$.
When counterfactual fairness does not hold, we quantify the total disparity by summing the individual differences:
\begin{equation} \label{EqCF}
\Delta CF(\pi, \pi') = \sum_{i \in \mathcal{N}} |J_i(\pi) - J_i(\pi')|
\end{equation}
\end{definition}


\begin{definition}[\textbf{Conditional Statistical Parity}] \label{Conditional Statistical Parity}
Let $LF \in (At \setminus \{z\})$ be a non-sensitive, legitimate attribute, whose value for agent $i$ is denoted by $LF_i$. Let $\mathcal{L}_{LF}$ be the set of possible values for $LF$. Conditional statistical parity is satisfied if demographic parity holds across all subgroups of agents that share the same value for the attribute $LF$.

Formally, for each value $v \in \mathcal{L}_{LF}$, let $\mathcal{N}_v = \{i \in \mathcal{N} \mid LF_i=v\}$. Conditional statistical parity holds if, for all $v \in \mathcal{L}_{LF}$:
$\bar{J}_{\mathcal{N}_1 \cap \mathcal{N}_v}(\pi) = \bar{J}_{\mathcal{N}_0 \cap \mathcal{N}_v}(\pi)$.

The total disparity is the sum of the disparities across each legitimate factor value:
\begin{equation} \label{EqCSP}
\Delta CSP(\pi, LF) = \sum_{v \in \mathcal{L}_{LF}} \left| \bar{J}_{\mathcal{N}_1 \cap \mathcal{N}_v}(\pi) - \bar{J}_{\mathcal{N}_0 \cap \mathcal{N}_v}(\pi) \right|
\end{equation}
\end{definition}



\subsection{Fairness Metrics for Fair PPO}
PPO focuses on maximising agents' rewards. This section extends PPO by incorporating fairness constraints directly into the optimisation process. Specifically, we add a penalty term to the PPO objective function, discouraging behaviours that diverge from any of the selected disparities above.
However, designing the penalty accounting only for collected rewards can limit the effectiveness of learning policies, particularly in stochastic environments and the early training process stage. Thus, we compose the penalty of two parts: a retrospective component based on the total rewards collected in the past, and a prospective component based on the critic's real-time value estimates.
We modify Eq.~\ref{LossFunc} using the $\Delta DP$ metric as follows: 
\begin{align*} 
L_t^{\text{Fair-PPO}}(\theta_i) = \hspace{-1.5cm} & \\
& \begin{aligned}
\hat{\mathbb{E}}_t \left[L_t^{\text{CLIP}}(\theta_i) - c_1 L_t^{\text{VF}}(\theta_i) + c_2 H[\pi_{\theta_i}](s_t) - \lambda \cdot L_{t}^{\text{fair}}\right]
\end{aligned}
\end{align*}
where $L_{t}^{\text{fair}}$ is one of the fairness penalties, and the hyperparameter $\lambda$ controls the magnitude of the fairness penalty. 

To formulate the group-based penalties, we first define the sample average return $\bar{G}$ and sample average value $\bar{V}$ for any agent subgroup $\mathcal{G} \subseteq \mathcal{N}$:
\begin{equation}
    \bar{G}_{\mathcal{G}}(\tau) = \frac{1}{|\mathcal{G}|} \sum_{i \in \mathcal{G}} G_i(\tau) \quad \text{and} \quad \bar{V}_{\mathcal{G}}(s_t) = \frac{1}{|\mathcal{G}|} \sum_{i \in \mathcal{G}} V_{\theta_i}(s_t)
\end{equation}

\paragraph{Demographic Parity Penalty.}
This penalty is the sample-based equivalent of the $\Delta DP$ metric, penalising the difference in average outcomes between the groups $\mathcal{N}_1$ and $\mathcal{N}_0$.
\begin{align} \label{PenaltyDP}
L_{t}^{\text{fair-DP}} = \alpha \cdot \left| \bar{G}_{\mathcal{N}_1}(\tau) - \bar{G}_{\mathcal{N}_0}(\tau) \right| + \beta \cdot \left| \bar{V}_{\mathcal{N}_1}(s_t) - \bar{V}_{\mathcal{N}_0}(s_t) \right|
\end{align}
where the first term, the retrospective component (weighted by $\alpha$), compares the average total return of the two groups over the last trajectory $\tau$. The second term, the prospective component (weighted by $\beta$), compares the groups' average value estimates for the current state $s_t$. The parameters $\alpha$ and $\beta$ balance their contributions.

\section{Experimental Setup}~\label{Experiments}
We test Fair-PPO in two MAS: a version of the Allelopathic Harvest (AH)~\cite{Leibo2019} and HospitalSim (HS), an original hospital simulation.\footnote{More details of the game in the supplementary material.} 
\paragraph{Allelopathic Harvest.}
In this setup, two groups of agents with distinct preferences, one for red berries and the other for blue berries, move in a grid and engage in cooperative dynamics within their respective groups, i.e., they plant and ripen berry bushes of their favourite colour and compete against the opposing group by blocking agents with opposing preferences. The objective for each group is to ensure the proliferation of their preferred berry, thereby maximising their rewards. The allelopathic element of AH is reflected in the bushes' spontaneous growth, which follows a linear function based on the number of red and blue bushes already on the ground. Thus, actions such as changing the colour of berries on a bush play a fundamental role in enhancing the growth of favoured bushes and ensuring future rewards for the group. 
Within each group, half of the agents can move every turn, while others are limited to moving only every two turns. This difference in mobility is a sensitive attribute which can be interpreted as an impairment.

\paragraph{HospitalSim.}
HospitalSim (HS) simulates a hospital where three agents, an escort dispatcher, a triage router, and a doctor manager, coordinate to maximise patient care efficiency while ensuring treatment fairness between sensitive and non-sensitive patients. 

The triage router assigns patients to six wards based on a multi-level mapping of symptoms/wards, and the expected wait time for each ward. Thus, the agent should balance a trade-off between sending patients to their specialised ward, which may be busy, versus a faster backup ward that offers a reduced reward. 
The escort dispatcher assigns available nurses and robots to impaired patients who need help moving in the hospital. Such an assignment is based on patients' priority, how long they have been waiting for an escort and the proportion of nurses and robots available. 
The doctor manager mitigates overcrowding by assigning swing doctors to different wards. This decision considers each ward's queue length, doctor availability, and the average patient priority.

\begin{figure}
    \centering
    \includegraphics[width=0.9\linewidth]{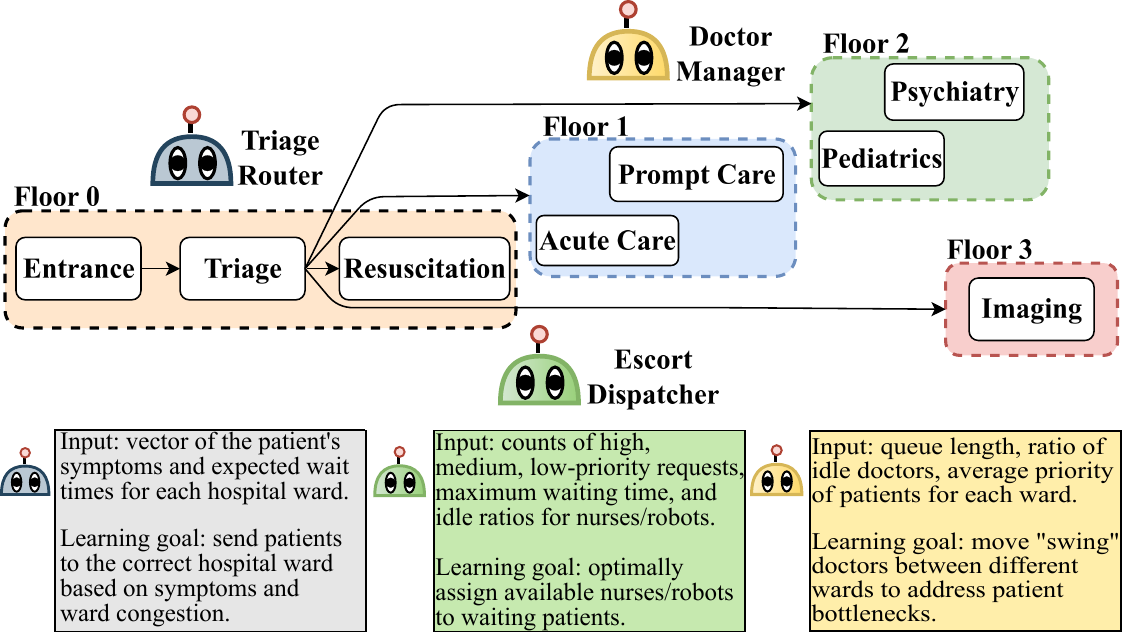}
    \caption{HospitalSim workflow and input/task descriptions for each agent.}
    \label{HS-diagram}
\end{figure}

Patients arrive throughout the day with defined priorities, illnesses, and physical impairments (sensitive attribute); impaired patients have slow movement and need assistance, leading to potential delays. The waiting time is penalised based on the priority of the patient. The system's learning agents integrate with rule-based agents, handling static functions like patient intake and diagnosis. A representation of HS is shown in Figure~\ref{HS-diagram}.

\subsection{Train}
In AH, we train two separate Fair-PPO models: one for agents with the sensitive attribute and another for those without. Training is conducted over $1000$ episodes, each representing a new game, with $3000$ time steps per episode. 
In HS, we train one model for each agent: the escort dispatcher, triage router, and doctor manager. Training spans $2000$ episodes, with each episode simulating a new hospital day of twelve hours and a stream of $300$ patients per day drawn from a generated pool. The simulation is event-driven, with parallel agents' actions that sequentially drive the game (thus, there are no time steps). Therefore, Fair-PPO employs a decentralised training approach, with a centralised fairness penalty calculation to add to the PPO objective function. 

In both games, after each episode, two penalty fairness components are calculated: the reward disparity is the difference in the total rewards accumulated by the non-sensitive group versus the sensitive group; the state-value disparity is the difference in the predicted state values between the two groups. Both components are normalised and weighted by $\alpha$ and $\beta$ parameters. However, while in AH the fairness penalty is calculated based on the rewards experienced by the learning agents (red and blue agents), in HS the penalty is based on the total and expected rewards of the patients who are not the learning agents.

We train Fair-PPO using a penalty weighted by $\alpha$ and $\beta$, addressing demographic parity, counterfactual fairness, and conditional statistical parity separately. The parameters $\alpha$ and $\beta$ range from 0 to 1, taking discrete values with a step of $0.25$, for a total of $25$ combinations. PPO is the specific instance of Fair-PPO with $\alpha = \beta = 0$.

\subsection{Test and evaluation}
We tested the policy on $500$ episodes of $1500$ time steps for AH and $500$ episodes (days) of twelve hours with a stream of $300$ patients per day drawn from a generated pool for HS (no time steps as explained in the previous section). Each metric is averaged over the time steps and episodes.
In both AH and HS, demographic parity is the difference between the rewards collected by agents with and without a sensitive attribute (impairment). For conditional statistical parity, the legitimate factor is agents' preference for red or blue berries in AH, and patients' priority (high, medium and low) in HS.

\subsection{Benchmarks}
We benchmark Fair-PPO against PPO ($\alpha=0$ and $\beta = 0$), FEN~\cite{Jiang2019} and SOTO~\cite{Zimmer2021}.
The other hyperparameters of each algorithm are reported in the supplementary material.

In both AH and HS, FEN is implemented following a hierarchical policy structure where a controller selects from several sub-policies. One sub-policy optimises for efficiency while the others explore fair strategies. For example, in AH, an efficient action is eating a preferred berry, while a fair one is ripening a berry for another agent. In HS, the triage router's efficiency policy learns to send a patient to the correct and least congested ward. In contrast, a fairness-oriented policy might prioritise patients with an impairment to better balance the utility across the entire system.
The agent can change the policy every $T_{macro}$ macro-timesteps. The controller learns to maximise a fair-efficient reward that encourages actions oriented to the utility of all agents while penalising any deviation from fairness. In AH, FEN shares the parameters among all agents, employing a gossip protocol to average utility, while in HS, three independent FEN instances are employed for each learning agent.

In AH and HS, SOTO is implemented by providing each learning agent with a dual PPO-based policy that balances individual efficiency with collective fairness. In both cases, a self-oriented stream maximises local rewards. In contrast, a team-oriented stream optimises a social welfare function using an $\alpha\text{-fairness}$ mechanism that weights an agent's advantage based on the utilities of others. The core difference of SOTO implementation in AH and HS is that in AH, the coordination of agents happens within each group (red and blue agents), and the fairness is calculated between non-sensitive and sensitive agents; in HS, the learning agents maximise their utility through the self-oriented policy, but the fairness-oriented policy minimises disparity between patients with and without impairment that are not learning agents.

\section{Results}~\label{Results}
In this section, we analyse the main findings from experiments in AH and HS. 

\paragraph{Fairness and efficiency measures.}
Our focus is on demographic parity and conditional statistical parity calculated on agents' rewards as the core fairness metrics. In AH, we also quantify fairness using the Gini index to assess reward distribution among agents, while overall efficiency is measured by the agents' average accumulated rewards throughout the simulation. For HS, we evaluate hospital efficiency using several key metrics: the average number of patients treated daily, patients' average waiting time for an escort, and the escorts' average travel time. We specifically evaluate the triage router policy by analysing the percentage of patients directed to the primary ward, backup ward, or misclassified. The doctor manager is assessed by the average number of trips made by swing doctors.

\subsection{Fair-PPO generates fairer policies than PPO}
Fair-PPO, through different combinations of $\alpha$ and $\beta$, produces several policies with lower demographic disparity than PPO in both AH and HS. Figure~\ref{FairPPO-PPO} shows three clusters of strategies (from light pink to dark purple) obtainable through adjusting the retrospective and prospective components of Fair-PPO in AH. Higher demographic parity generally corresponds to higher Price of Fairness, computed as the reward percentage agents renounce in favour of fairness compared to PPO. A similar result can be observed in HS, see the first (patients treated for efficiency) and the second column (demographic disparity for fairness) of Table~\ref{HS-test-results}. A lower disparity is also registered within the subgroups of agents (conditional statistical disparity), in both AH and HS: for example, Table~\ref{HS-test-results-CSP} first column, shows the improvement within the priority groups of patients (high, medium and low priority).

\subsection{Fair-PPO: comparison with FEN and SOTO}
\begin{figure*}
    \centering
    \includegraphics[width=1\linewidth]{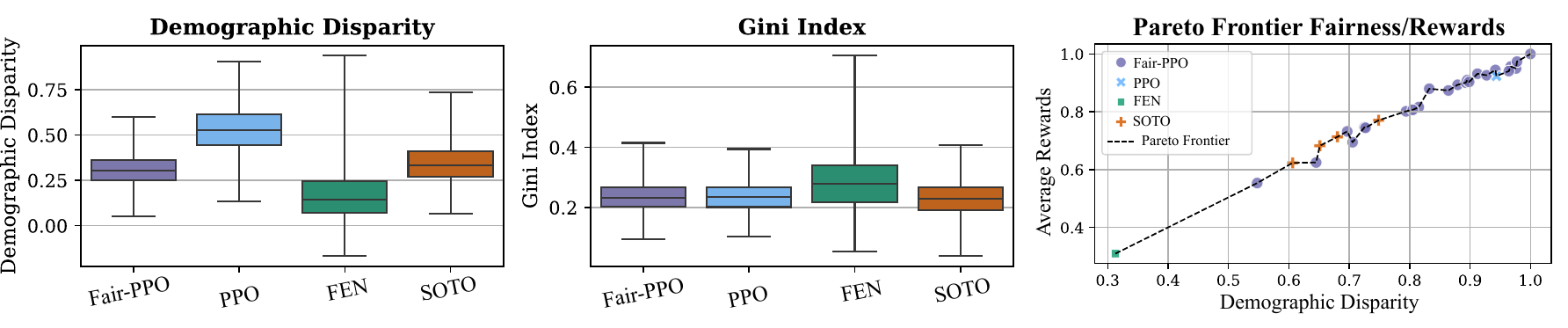}
    \caption{Allelopathic Harvest. Fair algorithms' performance (average demographic disparity and Gini index) for each architecture (Fair-PPO, PPO, FEN and SOTO) and disparity/rewards Pareto frontier plotted for all algorithms.}
    \label{DP-Pareto}
\end{figure*}

In AH, Fair-PPO matches SOTO on fairness-reward trade-offs, performing similarly in terms of equal reward distribution among agents. On the other hand, FEN records the lowest disparity but highly underperforms all algorithms in terms of inequality (Gini index) and average rewards. These results are reported in Figure~\ref{DP-Pareto}.

In HS, Fair-PPO finds the fairest solutions as reported in Table~\ref{HS-test-results}, revealing a trade-off between lower efficiency, in terms of daily average treated patients, and lower demographic disparity. On the other hand, SOTO and FEN produce very prolific policies in terms of daily average treated patients. Looking at Figure~\ref{HS-training-metrics}, where the training phase is reported, SOTO learns an efficient and fair policy, without succeeding in reaching the level of parity of Fair-PPO.  The same trend is confirmed by the test in Table~\ref{HS-test-results} and \ref{HS-test-results-CSP} throughout demographic disparity and conditional statistical disparity columns.

\subsection{Interpreting the strategies}
\begin{figure}
    \centering
    \includegraphics[width=1\linewidth]{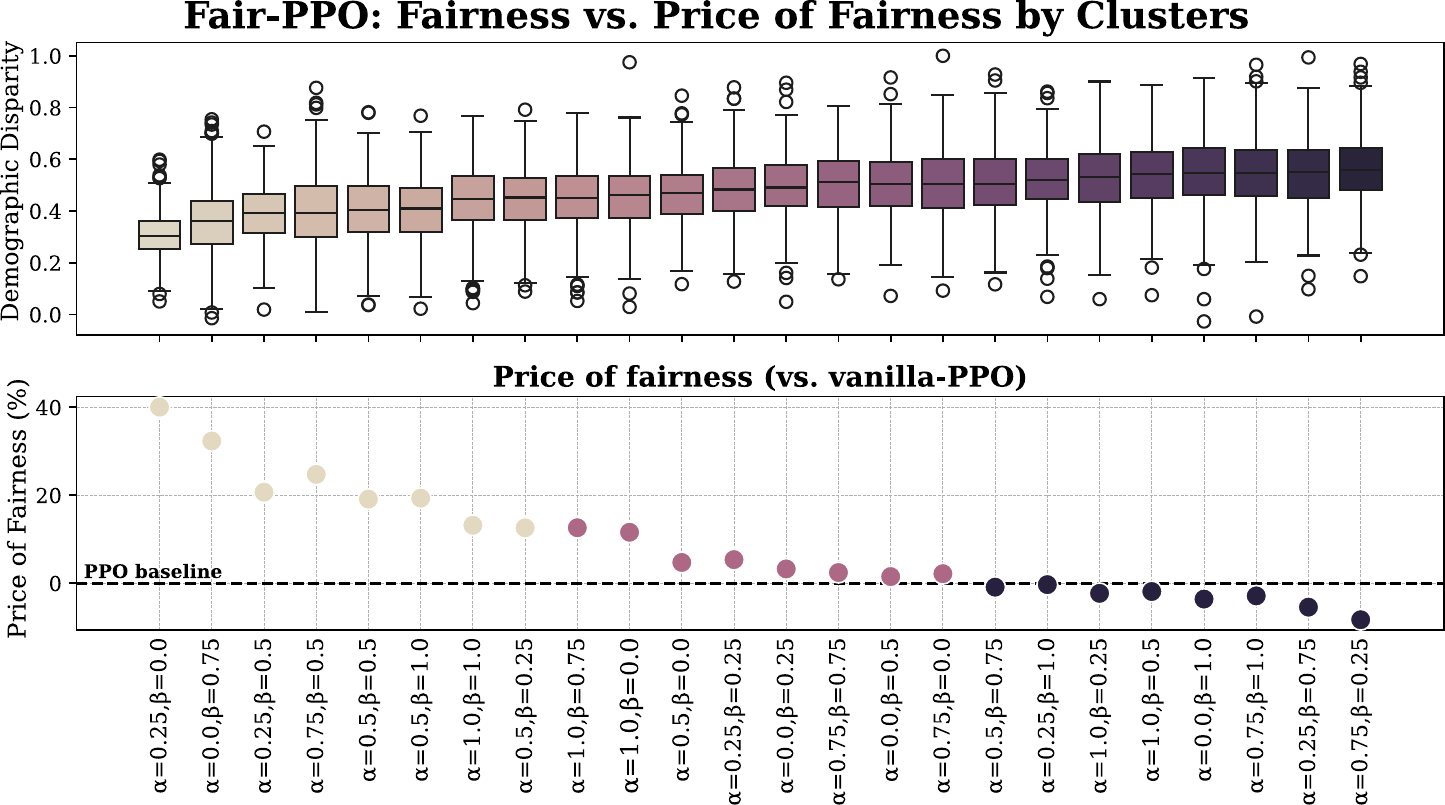}
    \caption{Allelopathic Harvest: Fair-PPO ($\alpha$, $\beta$) demographic disparity boxplots clusters (light pink, purple and dark purple) and the Price of Fairness vs. PPO.}
    \label{FairPPO-PPO}
\end{figure}

\begin{figure}
    \centering
    \includegraphics[width=1\linewidth]{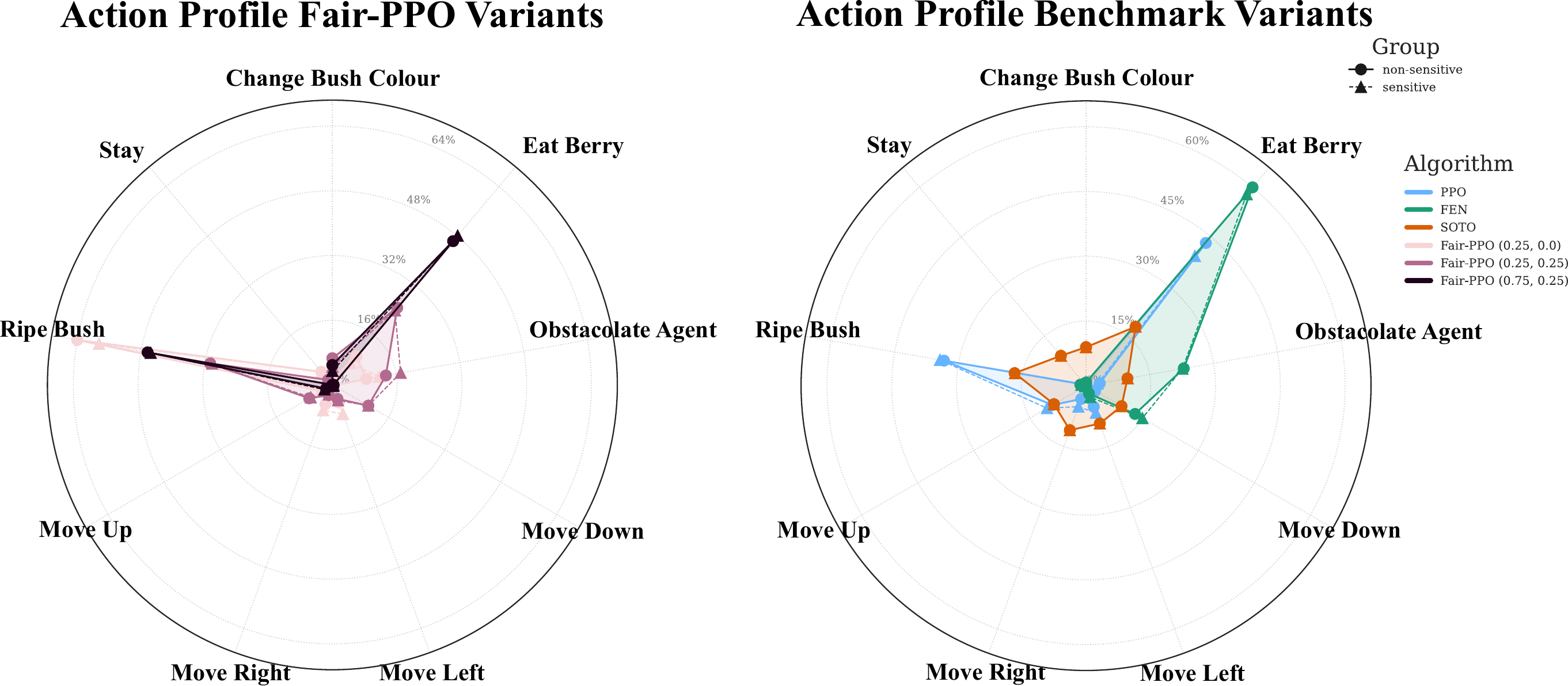}
    \caption{Allelopathic Harvest. Action frequencies of Fair-PPO and benchmarks. For Fair-PPO, we show one algorithm for each cluster (light pink, purple and dark purple), see Figure~\ref{FairPPO-PPO}.}
    \label{AH-Strategies}
\end{figure}

By varying $\alpha$ and $\beta$, Fair-PPO explores a wide variety of strategies in AH. For example, Figure~\ref{AH-Strategies}, left chart, shows the action frequencies of a representative of each cluster of Figure~\ref{FairPPO-PPO} (light pink, purple and dark purple).
Low demographic disparity (light pink) corresponds to a strategy that heavily prioritises ripening the bush, which is the action most beneficial for other agents collecting rewards. The highest disparity (dark purple) focuses more on berry consumption, the selfish action. SOTO reaches similar fairness results to Fair-PPO with lower disparity through a more uniformly distributed strategy, while FEN agents learns a berry-eating policy, which, later in the game, exhausts resources and dramatically decreases the rewards to collect.

\begin{figure}
    \centering
    \includegraphics[width=1\linewidth]{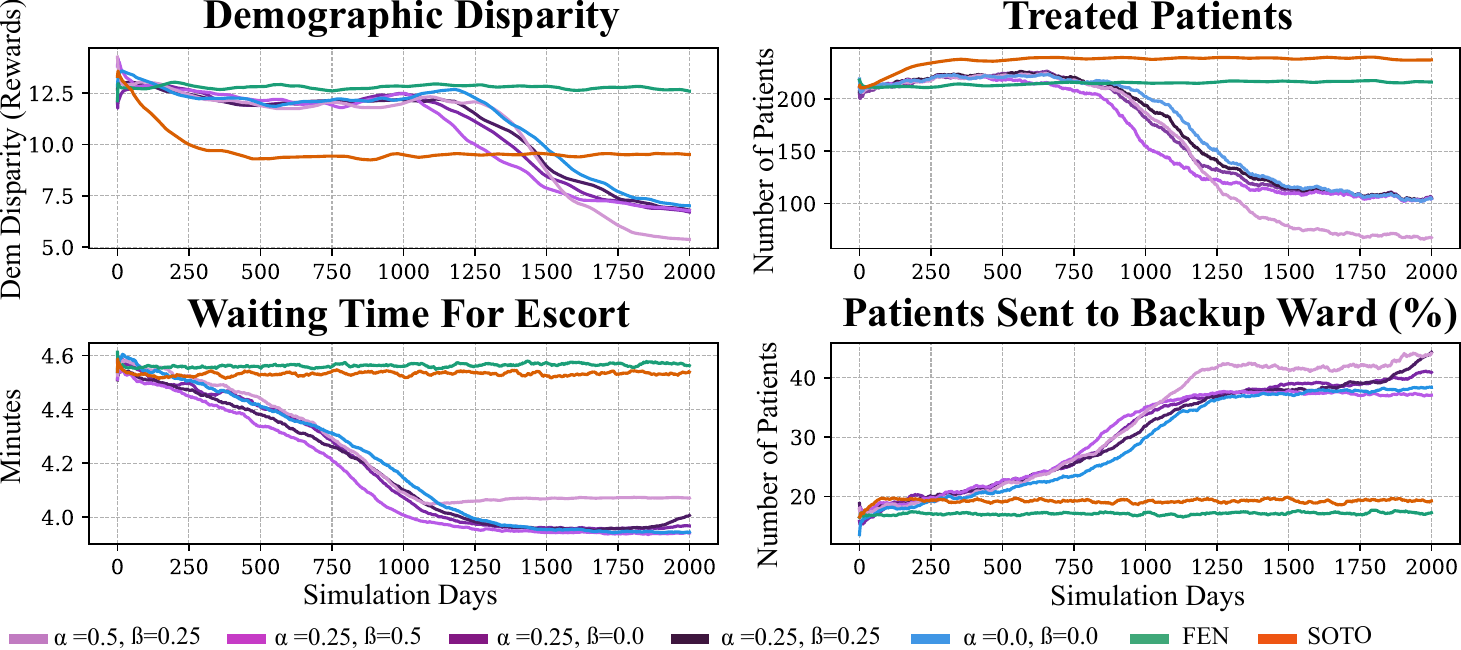}
    \caption{HospitalSim. Fairness (demographic disparity) and efficiency (daily treated patients and patients' waiting time for escort) metrics convergence over the training for Fair-PPO and benchmarks.}
    \label{HS-training-metrics}
\end{figure}

In HS, Fair-PPO reduces the disparity between impaired and unimpaired patients by alleviating overcrowding. Since the hospital must admit all 300 patients daily, Fair-PPO optimises the triage router to heavily direct patients to backup wards. This strategy is evidenced by the data in the last three columns of Table~\ref{HS-test-results} and further illustrated in the training graph in Figure~\ref{HS-training-metrics}. We also observe that a reduction in demographic disparity aligns with decreased patients' waiting times for escorts and lower escort travel time. Both metrics underscore improved access to supportive resources for impaired patients. In conclusion, although SOTO manages to optimise the triage router better than other algorithms to send the patients to the specialised ward and also to reduce the swing doctor moves (see Table~\ref{HS-test-results}, fifth column), it does not achieve the same parity level of Fair-PPO and PPO.

\begin{table*}
\centering
\sisetup{table-format = 4.2}
\resizebox{\textwidth}{!}{%
\begin{tabular}{
 l
 S[table-format=3.2]
 S[table-format=2.2]
 S[table-format=1.2]
 S[table-format=1.2]
 S[table-format=4.2]
 S[table-format=4.1]
 S[table-format=2.2]
 S[table-format=2.2]
}
\toprule
\textbf{Algorithm} & {\makecell{\textbf{Patients} \\ \textbf{Treated} \\ \textbf{(Daily Average)}}} & {\makecell{\textbf{Dem} \\ \textbf{Disparity} \\ \textbf{(Rewards)}}} & {\makecell{\textbf{Patient} \\ \textbf{Wait Escort} \\ \textbf{(Minutes)}}} & {\makecell{\textbf{Escort} \\ \textbf{Travel Time} \\ \textbf{(Minutes)}}} & {\makecell{\textbf{Swing Doctor} \\ \textbf{Moves} \\ \textbf{(Average)}}} & {\makecell{\textbf{Perfect Routing} \\ (\textbf{\% of Patients)}}} & {\makecell{\textbf{Backup Routing} \\ \textbf{(\% of Patients)}}} & {\makecell{\textbf{Incorrect Routing} \\ \textbf{(\% of Patients)}}} \\
\midrule
Fair-PPO $\alpha=0.25, \beta=0.25$  & 105.74 & 6.98 & 4.01 & 3.67 & 1923.93 & 6.04 & 44.93 & 49.03 \\
Fair-PPO $\alpha=0.25, \beta=0.5$  & 105.84 & 6.62 & 3.94 & 3.55 & 1925.52 & 6.00 & 37.13 & 56.86 \\
Fair-PPO $\alpha=0.5, \beta=0.25$  & 68.69  & 5.31 & 4.07 & 3.77 & 1924.72 & 3.83 & 43.78 & 52.39 \\
\midrule
PPO  & 105.92 & 6.83 & 3.94 & 3.56 & 1923.99 & 6.01 & 38.17 & 55.82 \\
\midrule
FEN & 223.17 & 11.76  & 4.54 & 4.09 & 1924.25  & 19.65  & 18.42  & 61.93 \\
\midrule
SOTO $\alpha\text{-fair}=1.0$  & 240.03 & 9.10 & 4.51 & 4.06 & 1167.36 & 27.64 & 21.17 & 51.19 \\
\bottomrule
\end{tabular}}%
\caption{HospitalSim. Average efficiency (patients treated) and fairness performance (demographic disparity) across Fair-PPO and benchmark, sorted according to the lowest demographic disparity. Further metrics are reported to complete the analysis.}
\label{HS-test-results}
\end{table*}

\begin{table}
\centering
\resizebox{0.5\textwidth}{!}{%
\begin{tabular}{
  l
  S[table-format=1.2]
  S[table-format=1.2]
  S[table-format=1.2]
  c 
}
\toprule
& \multicolumn{3}{c}{\textbf{Cond Stat Disparity}} & \textbf{Patients Treated} \\
\cmidrule(lr){2-4} \cmidrule(lr){5-5}
\textbf{Algorithm} & {\textbf{High}} & {\textbf{Medium}} & {\textbf{Low}} & (\textbf{Daily Average}) \\
\midrule
Fair-PPO $\alpha=1.0$ $\beta=0.5$ & 5.43 & 5.45 & 7.63 & 63.75 \\ 
Fair-PPO $\alpha=0.25$ $\beta=1.0$ & 5.43 & 5.43 & 7.54 & 64.05 \\ 
Fair-PPO $\alpha=1.0$ $\beta=0.0$ & 5.19 & 5.39 & 7.44 & 64.15 \\ 
\midrule
PPO & 5.32 & 5.30 & 8.10 & 63.85 \\
\midrule
FEN & 11.77 & 11.80 & 11.63 & 223.76 \\
\midrule
SOTO $\alpha\text{-fair}=2.0$ & 9.17 & 9.14 & 9.11 & 239.95 \\
\bottomrule
\end{tabular}%
}
\caption{HospitalSim. Average conditional statistical disparity (high, medium, low patients' priority) and efficiency (patients treated) performance across Fair-PPO and benchmark, sorted for best demographic disparity.}
\label{HS-test-results-CSP}
\end{table}

\section{Conclusion}
This paper extends PPO by incorporating a penalty term based on fairness metric violations in the objective function. We design two penalty components: a retrospective component that addresses fairness violations based on past rewards and a prospective component that anticipates future fairness violations by leveraging the value function to estimate upcoming rewards. We refer to this algorithm as Fair-PPO. 
We found that Fair-PPO can discover various strategies to reduce unfairness across metrics for different levels of efficiency in both cooperative and competitive games (AH) and coordination-based simulations (HS). 
This work represents a first step towards a fairness-aware PPO based on metrics in MAS involving agents with and without sensitive attributes. 
The implementation of the HS aims to stimulate fair reinforcement learning research towards real-world applications.   

\section*{Ethical Statement}
This work shows how a penalty-based approach can make a multi-agent policy converge towards fairer behaviours. The penalty is based on metrics involving agents with and without sensitive attributes. 
Thus, this work requires careful ethical considerations. First, the fairness metrics used and the sensitive attribute are an inevitable simplification of fairness in reality. In particular, we recognise that the sensitive attribute cannot capture the complexity of real-world scenarios. Second, our results show a trade-off between fairness and efficiency of a system. The deployment of a model requires transparency and awareness in the presence of this trade-off, in particular when the applications are intended to imitate a hospital, as in the HS. In conclusion, we stress that the findings of the paper should be only a stimulus for discussion on computational simulations and fairness, rather than a direct basis for policy.

\bibliography{AnonymousSubmission/LaTeX/arxiv_bibliography}
\clearpage
\clearpage
\section*{Supplementary Material}
\section*{Fairness Metrics for Fair-PPO: Counterfactual Fairness and Conditional Statistical Parity}
In this section, we formalise the incorporation of counterfactual fairness and conditional statistical parity in PPO loss. This section should be intended as complementary to the subsection ``Fairness Metrics for Fair-PPO'' of the main paper.

\paragraph{Counterfactual Fairness Penalty.}
This penalty minimises the change in outcomes for each agent when its sensitive attribute is notionally flipped.
\begin{align} \label{PenaltyCF}
L_{t}^{\text{fair-CF}} = \alpha \cdot \sum_{i \in \mathcal{N}} \left| G_i(\tau) - G_i^{\prime}(\tau) \right| + \beta \cdot \sum_{i \in \mathcal{N}} \left| V_{\theta_i}(s_t) - V_{\theta_i}^{\prime}(s_t) \right|
\end{align}
where $G_i(\tau)$ and $V_{\theta_i}(s_t)$ are the return and value estimate for agent $i$ in the factual system, while $G_i^{\prime}(\tau)$ and $V_{\theta_i}^{\prime}(s_t)$ are their equivalents in the counterfactual system.

\paragraph{Conditional Statistical Parity Penalty.}
This penalty is the sample-based equivalent of the $\Delta CSP$ metric. It minimises demographic parity disparities conditioned on a legitimate, non-sensitive attribute $LF \in At \setminus \{z\}$.
\begin{align} \label{PenaltyCSP}
L_{t}^{\text{fair-CSP}} = & \alpha \cdot \sum_{v \in \mathcal{L}_{LF}} \left| \bar{G}_{\mathcal{N}_1 \cap \mathcal{N}_v}(\tau) - \bar{G}_{\mathcal{N}_0 \cap \mathcal{N}_v}(\tau) \right| \nonumber \\
& + \beta \cdot \sum_{v \in \mathcal{L}_{LF}} \left| \bar{V}_{\mathcal{N}_1 \cap \mathcal{N}_v}(s_t) - \bar{V}_{\mathcal{N}_0 \cap \mathcal{N}_v}(s_t) \right|
\end{align}
This sums the outcome disparities between groups $\mathcal{N}_1$ and $\mathcal{N}_0$ across each subgroup of agents ($\mathcal{N}_v$) that share the same value $v$ for the legitimate attribute $LF$.

\section*{The Allelopathic Harvest and HospitalSim}
In this section, we provide the details of the implementation of the Allelopathic Harvest (AH) and HospitalSim (HS).
\paragraph{Allelopathic Harvest.}
The AH environment is a 2D grid world with 40 agents and 60 berry bushes. 
Half of the players favour red berries and the other blue. Agents can move in the grid (up, down, right, left), eat berries, change the colour of the berries on a bush, plant and ripen berry bushes of their favourite colour and block, for one turn, agents with opposed preferences. This spectrum of actions makes the AH a cooperative and competitive game.  

Within each group, half of the agents can move every turn, while others are limited to moving only every two turns. This difference in mobility is a sensitive attribute which can be interpreted as an impairment.

The bushes in the grid spontaneously grow following a linear function based on the number of red and blue bushes already on the ground. Furthermore, each bush has a lifespan at the end of which it dies. The berries also regrow on the bushes. We report the details of the AH in Table~\ref{Baseline-AH}, which is the baseline of the experiments.

\paragraph{Reward system in the AH.}
The AH rewards an agent through different actions directed at the berry bushes or other agents. Berry consumption is positively rewarded, in particular if the berry consumed matches the agent's preferences. Preparing a berry for consumption through the ripening action is rewarding, in particular if the berry has the colour preferred by the ripening agent. Also, an agent is rewarded for changing the colour of the berries produced by a bush, especially if the berries previously produced were of a different colour. Lastly, an agent is rewarded by obstructing another agent, in particular if that agent has a different berry preference.

\begin{table}[h!]
\centering
\small
\begin{tabular}{|l|l|}
\hline
\textbf{Variable} & \textbf{Baseline} \\ \hline
Initial number of agents & 40 \\ \hline
Initial number of bushes & 30 \\ \hline
Red/blue agent \%  & 50\%\\ \hline
Berry regrowth    & 3ts \\ \hline
Bush lifespan & 120ts\\ \hline
Bush growth rate  & 2ts\\ \hline

\end{tabular}
\caption{Setup of the Allelopathic Harvest (ts = timesteps).}
\label{Baseline-AH}
\end{table}

\paragraph{HospitalSim.}
HS simulates the operations of a hospital where three learning agents, a triage router, an escort dispatcher, and a doctor manager work to optimise patient care. In addition to them, clerks, nurses, robots, doctors and patients act in the hospital as non-learning agents (they do not have a learning policy). The HS is then a system where learning agents integrate with non-learning agents.

The hospital consists of an entrance, a triage room, six wards to treat patients, corridors connecting the rooms positioned at different distances and floors and an exit. Any agent who moves is assigned a speed, so that it is possible to compute the time to cover a distance from one room to another.

\paragraph{The patients.}
Each patient arrives with a priority level (high, medium, low), an impairment level (none, low, high) and an illness. The impairment level is associated with the speed of the patient (75 for none, 60 for low, and 45 for high), and only patients without impairment can move by themselves in the hospital, while the others need assistance (from a nurse or a robot). The speed of the nurse or robot (who have respective speeds of 90 and 100) adapts to the patient they are assigned to.

Patients arrive with a mix of priorities, impairments, and illnesses. The priorities are uniformly distributed ($33.3\%$ to have one of the priorities) as well as the illnesses ($16.7\%$ chances of having an illness that should be treated in one of the six wards). On the other hand, a patient has a $60\%$ chance of having no impairment, a $25\%$ chance of having low impairment and $15\%$ chance of having high impairment. 
Each illness is assigned to a series of symptoms as reported in Table~\ref{Illness-symptoms}. The symptoms are the classification proxies used by the triage router to decide the destination of the patient.

During a day (episode), patients arrive according to a Poisson arrival distribution, based on an arrival rate and time. The arrival rate can be high for the peak or low for the other times. The arrival time is instead random, but the likelihood of a patient showing up changes depending on whether it is peak or off-peak time. 

\begin{table}[h!]
\centering
\begin{tabular}{|l|p{5.5cm}|}
\hline
\textbf{Illness} & \textbf{Symptoms} \\
\hline
Pediatric & Is a child, fever, cough \\
\hline
General & Fever, minor pain \\
\hline
Cardio & Chest pain, shortness of breath, high blood pressure \\
\hline
X-ray & Suspected fracture, minor pain \\
\hline
Psychiatric & Confusion, high blood pressure \\
\hline
Emergency & Unconsciousness, chest pain, shortness of breath \\
\hline
\end{tabular}
\caption{Mapping of Illnesses to Symptoms.}
\label{Illness-symptoms}
\end{table}

\paragraph{The learning agents.}
The learning agents are three: a triage router, an escort dispatcher and a doctor manager. Each agent is programmed to work in coordination with the system, i.e., its input hides a mix of subtasks. 

The triage router decides which ward a patient should be assigned to. Such an assignment is based on: (i) a multi-level mapping of symptoms/wards that the triage router should learn to classify patients correctly, and (ii) the expected wait time for each ward. Thus, the agent is not only a classification but should learn to optimise based on the waiting times in different wards. Also, most illnesses have a dedicated specialised ward for treatment. If patient load becomes too high, backup wards are available where the illness can still be treated, though with a reduced (weighted) reward to reflect it being a less specialised option. For an illness-ward mapping and the reward weights, see Table~\ref{Illness-ward}.

\begin{table}[h!]
\centering
\begin{tabular}{|l|p{5.0cm}|}
\hline
\textbf{Illness} & \textbf{Main/Backup Wards} \\
\hline
Pediatric & Pediatrics: $1.0$, PromptCare: $0.6$ \\
\hline
General & PromptCare: $1.0$, AcuteCare: $0.5$ \\
\hline
Cardio & AcuteCare: $1.0$,  Resuscitation: $0.7$, PromptCare: $0.4$ \\
\hline
X-ray & Imaging: $1.0$, PromptCare: $0.5$ \\
\hline
Psychiatric & Psychiatry: $1.0$, PromptCare: $0.2$ \\
\hline
Emergency & Resuscitation: $1.0$ \\
\hline
\end{tabular}
\caption{Mapping of illnesses to their primary and backup wards, including relative reward weights for patients treated in backup facilities.}
\label{Illness-ward}
\end{table}

The escort dispatcher assigns available nurses and robots to patients if they need help to move from one area to another in the hospital. Such an assignment is based on: (i) the number of high, medium, and low priority (patients) requests, (ii) the longest time any patient has been waiting for an escort, (iii) the proportion of nurses and robots who are currently idle.

The doctor manager assigns a number of swing doctors to the different wards to handle overcrowding. Such an assignment is based on: (i) the number of patients waiting in that ward, (ii) the proportion of doctors in that ward who are currently available, (iii) the average priority of patients waiting in the queue, giving a sense of the urgency in that ward.
    
We report the details of the HS in Table~\ref{Baseline-HS}, which is the baseline of the experiments.

\paragraph{Reward system in the HS.} The HS rewards timely patient care while penalising delays (inefficient escort) and poor assignment. Patients are rewarded based on their satisfaction, positively for advancing in their care pathway, and negatively if they need to wait too long in the different hospital areas, considering their priority level. Escort dispatchers and doctor managers are rewarded for promptly assigning skilled (nurses instead of robots) and nearby staff to high-priority patients and for correctly routing individuals to the right wards. Conversely, they are penalised for leaving patients waiting and for poor assignment of nurses, robots and doctors, penalising inefficiency.

\begin{table}[h!]
\centering
\small
\begin{tabular}{|l|l|}
\hline
\textbf{Variable} & \textbf{Baseline} \\ \hline
Num clerks (entrance) & 30 \\ \hline
Num nurses (total) & 60 \\ \hline
Num robots (total) & 30 \\ \hline
Num triage dispatcher (total) & 30 \\ \hline
Num swing doctors (total)  & 18  \\ \hline
Num ward doctors (per ward)  & 10  \\ \hline
Num treatment wards  & 6  \\ \hline
\end{tabular}
\caption{Setup of the HospitalSim.}
\label{Baseline-HS}
\end{table}

\section*{Fair-PPO Details in the AH and HS}
In this section, we provide the details of the implementation of Fair-PPO in the AH and HS. 
\paragraph{AH.} Fair-PPO initialise two distinct PPO instances, one for sensitive and one for non-sensitive agents, implementing parameters sharing within these two groups. The implementation employs a decentralised learning (each group of agents learns from its own experience) with a centralised fairness coordination. 

After each episode, two penalty fairness components are calculated: the reward disparity is the difference in the total rewards accumulated by the non-sensitive group versus the sensitive group; the state-value disparity is the difference in the predicted state values between the two groups. Both components are normalised and weighted by $\alpha$ and $\beta$ parameters. Finally, a dynamic $\lambda$  is calculated to balance the magnitude of the fairness penalty with the standard PPO loss.

\paragraph{HS.} Fair-PPO initialise three distinct PPO instances, respectively for the escort dispatcher, the triage router and the doctor manager, which are responsible for different aspects of hospital operations and act independently (decentralised action). The fairness penalty is calculated at the end of each episode and thus centralised. However, unlikely the AH, the fairness penalty is calculated based on the rewards experienced by the patients, divided into sensitive and non-sensitive groups (with and without impairment), who are not the learning agents. The two penalty components are then based on the total and the predicted rewards of the patients obtained through the actions of the three learning agents. The parameters $\alpha$, $\beta$ and $\lambda$ work in the same way as in the AH.   

\section*{Benchmark Details: FEN and SOTO in the AH and HS}

In this section, we provide the details of the implementation of FEN and SOTO in the Allelopathic Harvest (AH) and HospitalSim (HS). 

\subsection{FEN Details}
\paragraph{AH.} In the AH, FEN is implemented as a hierarchical reinforcement learning algorithm designed to provide agents with an efficient policy to collect berries from bushes and promote fairness. A single FEN algorithm shared the set of parameters to control all 40 agents. The hierarchical architecture of FEN is composed of a controller that learns to select from a set of lower-level sub-policies. The controller makes a macro-decision, and the chosen sub-policy then interacts with the environment for a set number of steps ($T_{\text{macro}}$).

More precisely, the controller is trained, through Proximal Policy Optimisation (PPO), to understand through a reward-based feedback, if a chosen sub-policy performs well in terms of efficiency and fairness. If the reward is high (both satisfactory efficiency and fairness), then the policy is assigned a higher probability of being chosen and vice versa.   

In the AH, two sub-policies are trained with PPO: an efficiency-oriented and a fairness-oriented policy. The efficiency policy purely maximises the rewards of agents. For example, eating berries can be a purely efficient policy as the reward associated with that action is maximised. The fairness policy is incentivised to explore different, potentially fair, behaviours. For example, although associated with a lower reward, the ripe-based policy can make berries available to agents with lower rewards and incentivise reward equality.

We report the hyperparameters used to train FEN in the AH in Table~\ref{Hyperparams-FEN-AH}.

\begin{table}[h!]
\centering
\small
\begin{tabular}{|l|l|}
\hline
\textbf{Hyperparameter} & \textbf{Value} \\
\hline
Learning Rate (lr) & $1 \times 10^{-4}$ \\
\hline
Discount Factor ($\gamma$) & 0.99 \\
\hline
PPO Epsilon Clip & 0.1 \\
\hline
PPO Update Epochs ($k_{epochs}$) & 5 \\
\hline
Batch Size & 256 \\
\hline
Entropy Coefficient & 0.01 \\
\hline
Value Loss Coefficient & 0.5 \\
\hline
Number of Sub-policies ($k_{sub}$) & 2 \\
\hline
Macro-step Size ($T_{macro}$) & 10 \\
\hline
Reward Scale ($c$) & 1.0 \\
\hline
Fairness Epsilon Denominator ($\epsilon$) & $1 \times 10^{-6}$ \\
\hline
\end{tabular}
\caption{Hyperparameters to train FEN in the AH.}
\label{Hyperparams-FEN-AH}
\end{table}

\paragraph{HS.}
In the HS, FEN is designed to provide three learning agents, the escort dispatcher, triage router, and doctor manager, with an efficient and fair policy to manage patient flow. Each agent is governed by a FEN algorithm with same architecture and hyperparameters. A controller learns how to switch from the efficient to the fair sub-policies following the same logic we described for the AH. In the AH, four sub-policies are trained with PPO, one of which is designed to maximise efficiency. For example, the triage router is designed to assign a patient to a specific ward. The efficiency policy focuses on minimising treatment delays, while the other sub-policies are free to explore other strategies, such as prioritising patients who have been waiting longer, regardless of their formal priority level. In this way, the controller can select a policy to improve the reward gap between patients with and without impairment (used as a fairness score).

We report the hyperparameters used to train FEN in the AH in Table~\ref{Hyperparams-FEN-HS}.

\begin{table}[h!]
\centering
\small
\begin{tabular}{|l|l|}
\hline
\textbf{Hyperparameter} & \textbf{Value} \\
\hline
Learning Rate (lr) & $1 \times 10^{-5}$ \\
\hline
Discount Factor ($\gamma$) & 0.99 \\
\hline
PPO Epsilon Clip & 0.1 \\
\hline
PPO Update Epochs ($k_{\text{epochs}}$) & 10 \\
\hline
Batch Size & 64 \\
\hline
Entropy Coefficient & 0.01 \\
\hline
Value Loss Coefficient & 0.5 \\
\hline
Number of Sub-policies ($k_{\text{sub}}$) & 4 \\
\hline
Macro-step Size ($T_{\text{macro}}$) & 50 \\
\hline
Reward Scale ($c$) & 100.0 \\
\hline
Fairness Epsilon Denominator ($\epsilon$) & 0.1 \\
\hline
\end{tabular}
\caption{Hyperparameters to train FEN in the HS.}
\label{Hyperparams-FEN-HS}
\end{table}

\subsection{SOTO Details}
\paragraph{AH.} In the AH, SOTO is initialised for all 40 agents, relying on two sub-policies, one self and the other team-oriented, with the double objective of learning, through PPO, efficiency and fairness. The self-oriented policy purely maximises the agent's rewards, while the team-oriented policy aims at maximising a social welfare function by considering global environment features, e.g., the ratio of ripe blue and red bushes.  

The choice whether to follow the self or team-oriented policy is decided by the parameter $\beta-\text{proportion}$. At the beginning of the training, the chance of choosing the self-oriented policy is higher, while later in training, the agent is more likely to choose the team policy. The team policy is updated through tracking which agents collected lower rewards, and should be directed to select actions that improve them. An $\alpha\text{-fairness}$ hyperparameter controls how strongly the fairness is enforced. In the AH, the social welfare function is maximised for all agents, as the SOTO key concept is to automatically identify and prioritise agents that are performing poorly, regardless of their group partnership.

We report the hyperparameters used to train SOTO in the AH in Table~\ref{Hyperparams-SOTO-AH}.

\begin{table}[h!]
\centering
\small
\begin{tabular}{|l|l|}
\hline
\textbf{Hyperparameter} & \textbf{Value} \\
\hline
$\alpha\text{-fairness}$ & $0.9, 1.0, 2.0, 5.0$ \\
\hline
$\beta\text{-proportion}$ & 0.5 \\
\hline
Learning Rate (lr) & $1 \times 10^{-4}$ \\
\hline
Discount Factor ($\gamma$) & 0.99 \\
\hline
PPO Epsilon Clip & 0.2 \\
\hline
PPO Update Epochs ($k_{epochs}$) & 5 \\
\hline
Batch Size & 256 \\
\hline
Entropy Coefficient & 0.05 \\
\hline
\end{tabular}
\caption{Hyperparameters to train SOTO in the AH.}
\label{Hyperparams-SOTO-AH}
\end{table}

\paragraph{HS.} In the HS, SOTO is implemented to provide a policy to the three learning agents: the escort dispatcher, the triage router, and the doctor manager. SOTO architecture and hyperparameters are identical for each agent. SOTO implements a two-headed policy: one is self-oriented and aims to maximise the rewards of the agent, while the other, team-oriented, learns a fair policy based on a social welfare function. 

The choice of whether to follow one or the other policy is similar to the one described for the AH. The team policy is updated by tracking the disparity in outcomes between patient groups, in particular, the ratio of average rewards between non-impaired and impaired patients (who are non-learning agents in this environment). The ratio is used to weight the policy update, directing the agents to select actions that reduce this fairness gap.

We report the hyperparameters used to train SOTO in the AH in Table~\ref{Hyperparams-SOTO-HS}.

\begin{table}[h!]
\centering
\small
\begin{tabular}{|l|l|}
\hline
\textbf{Hyperparameter} & \textbf{Value} \\
\hline
$\alpha\text{-fairness}$ & $0.9$ \\
\hline
$\beta\text{-proportion}$ & 0.50 \\
\hline
Learning Rate (lr) & $5 \times 10^{-4}$ \\
\hline
Discount Factor ($\gamma$) & 0.99 \\
\hline
PPO Epsilon Clip & 0.2 \\
\hline
PPO Update Epochs ($k_{epochs}$) & 5 \\
\hline
Batch Size & 64 \\
\hline
Entropy Coefficient & 0.01 \\
\hline
\end{tabular}
\caption{Hyperparameters to train SOTO in the HospitalSim.}
\label{Hyperparams-SOTO-HS}
\end{table}

\section*{Additional Results}
In this section, we present the following additional results as a complement to the ones presented in the main paper.

In AH, for each group of agents (red and blue berry preference), we present the results (conditional statistical parity) of Fair-PPO and the benchmarks in Table~\ref{csp-red} and ~\ref{csp-blue}. The same results can be visualised in Figure~\ref{csp-red} and Figure~\ref{csp-blue}. Jain's Fairness Index (JFI) is used to quantify the equality of outcomes among a group of agents and produces a score bounded between $0$ and $1$, where $1$ represents perfect equality. The Normalised Nash Social Welfare (NNSW) evaluates both the efficiency and fairness of an outcome, with a strong emphasis on the well-being of the worst-off agents.
 
For the HS, we report the complete results of Tables 1 and 2 of the main paper in Table~\ref{HS-test-results-full} and \ref{HS-test-results-full-csp}. These tables show the results for Fair-PPO with all the combinations of $\alpha$ and $\beta$ (range from $0$ to $1$ with a step of $0.25$) and SOTO with all $\alpha\text{-fair}$ (respectively $0.9$, $1.0$, $2.0$ and $5.0$).

\begin{table}[h!]
\begin{tabular}{lcccc}
\toprule
Algorithm & Disparity (↓) & Gini (↓) & JFI (↑) & NNSW (↑) \\
\midrule
Fair-PPO & 0.182 & 0.178 & 0.874 & 0.914 \\
PPO & 0.315 & 0.182 & 0.871 & 0.914 \\
FEN & 0.068 & 0.219 & 0.824 & 0.859 \\
SOTO & 0.217 & 0.184 & 0.867 & 0.909 \\
\bottomrule
\end{tabular}
\caption{Mean metrics (best disparity for red berry preference agents group) per algorithm.}
\label{csp-red}
\end{table}

\begin{table}[h!]
\begin{tabular}{lcccc}
\toprule
Algorithm & Disparity (↓) & Gini (↓) & JFI (↑) & NNSW (↑) \\
\midrule
Fair-PPO & 0.159 & 0.165 & 0.886 & 0.924 \\
PPO & 0.313 & 0.177 & 0.875 & 0.916 \\
FEN & 0.105 & 0.275 & 0.758 & 0.769 \\
SOTO & 0.196 & 0.173 & 0.878 & 0.917 \\
\bottomrule
\end{tabular}
\caption{Mean metrics (best disparity for blue berry preference agents group) per algorithm.}
\label{csp-blue}
\end{table}

\begin{figure*}
    \centering
    \includegraphics[width=1\linewidth]{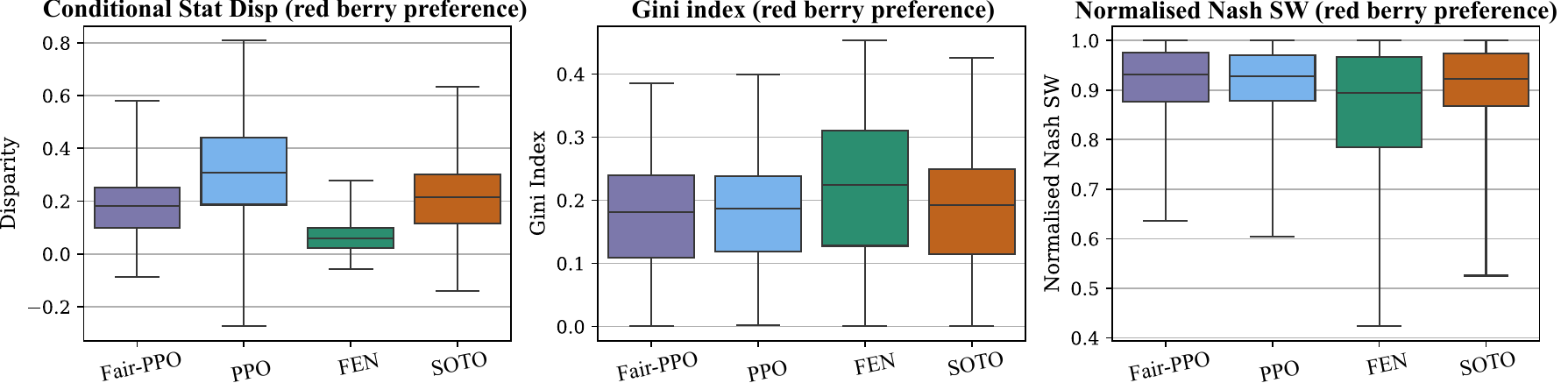}
    \caption{Allelopathic Harvest. Fairness (conditional statistical parity), Gini index and NNSW for Fair-PPO and benchmarks for the subgroups of agents with red berry preference. The box plots refer to the algorithms with the lowest disparity.}
    \label{csp-red}
\end{figure*}

\begin{figure*}
    \centering
    \includegraphics[width=1\linewidth]{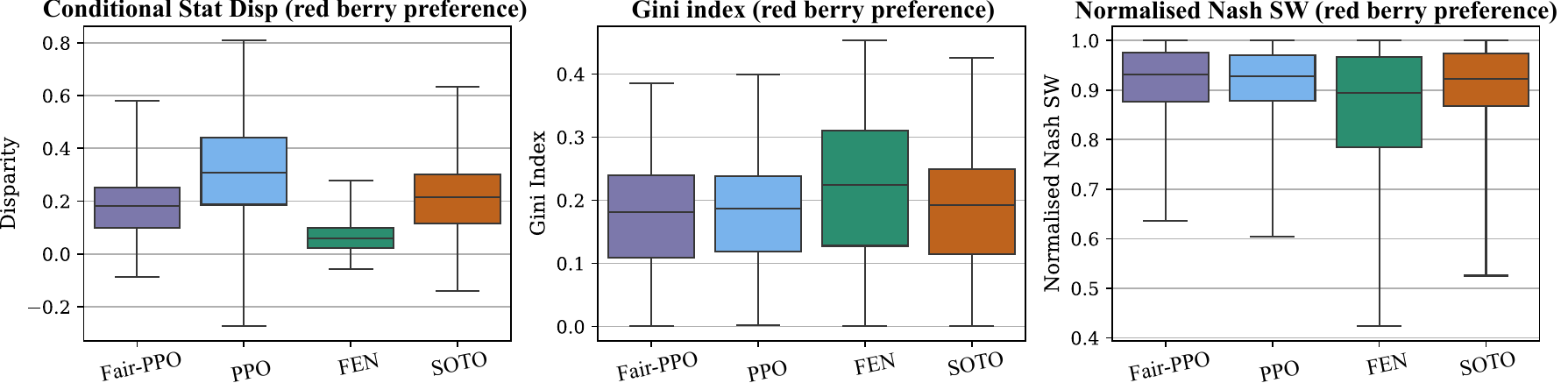}
    \caption{Allelopathic Harvest. Fairness (conditional statistical parity), Gini index and NNSW for Fair-PPO and benchmarks for the subgroups of agents with blue berry preference. The box plots refer to the algorithms with the lowest disparity.}
    \label{csp-blue}
\end{figure*}

\begin{table*}
\centering
\resizebox{\textwidth}{!}{%
\begin{tabular}{l c c c c c c c c}
\toprule
\textbf{Algorithm} & \textbf{Patients Treated} & \textbf{Dem Disparity} & \textbf{Patient Wait} & \textbf{Escort Travel} & \textbf{Swing Doctor} & \textbf{Perfect Routing} & \textbf{Backup Routing} & \textbf{Incorrect Routing} \\
 & \textbf{(Daily Average)} & \textbf{(Rewards)} & \textbf{(Minutes)} & \textbf{Time (Minutes)} & \textbf{Moves (Average)} & \textbf{(\% of Patients)} & \textbf{(\% of Patients)} & \textbf{(\% of Patients)} \\
\midrule
Fair-PPO $\alpha=0.0, \beta=0.25$ & 106.57 & 7.10 & 3.95 & 3.56 & 1923.70 & 6.04 & 38.03 & 55.93 \\
Fair-PPO $\alpha=0.0, \beta=0.5$ & 104.47 & 7.01 & 3.95 & 3.56 & 1923.45 & 5.91 & 39.05 & 55.03 \\
Fair-PPO $\alpha=0.0, \beta=0.75$ & 206.59 & 13.39 & 4.05 & 3.64 & 1925.62 & 13.89 & 33.89 & 52.22 \\
Fair-PPO $\alpha=0.0, \beta=1.0$ & 104.63 & 6.96 & 3.95 & 3.56 & 1923.20 & 5.93 & 39.11 & 54.96 \\
Fair-PPO $\alpha=0.25, \beta=0.0$ & 106.81 & 6.52 & 3.97 & 3.59 & 1924.73 & 6.09 & 40.89 & 53.02 \\
Fair-PPO $\alpha=0.25, \beta=0.25$ & 105.74 & 6.98 & 4.01 & 3.67 & 1923.93 & 6.05 & 44.93 & 49.03 \\
Fair-PPO $\alpha=0.25, \beta=0.5$ & 105.84 & 6.62 & 3.94 & 3.55 & 1925.52 & 6.01 & 37.13 & 56.86 \\
Fair-PPO $\alpha=0.25, \beta=0.75$ & 85.67 & 8.48 & 4.06 & 3.75 & 1924.22 & 4.85 & 43.62 & 51.53 \\
Fair-PPO $\alpha=0.25, \beta=1.0$ & 118.03 & 9.21 & 3.95 & 3.56 & 1924.34 & 6.77 & 37.49 & 55.74 \\
Fair-PPO $\alpha=0.5, \beta=0.0$ & 108.02 & 7.03 & 3.97 & 3.59 & 1924.03 & 6.15 & 40.45 & 53.40 \\
Fair-PPO $\alpha=0.5, \beta=0.25$ & 68.69 & 5.31 & 4.07 & 3.77 & 1924.72 & 3.83 & 43.78 & 52.39 \\
Fair-PPO $\alpha=0.5, \beta=0.5$ & 105.17 & 7.10 & 3.95 & 3.56 & 1925.35 & 5.95 & 39.04 & 55.01 \\
Fair-PPO $\alpha=0.5, \beta=0.75$ & 108.05 & 7.29 & 3.95 & 3.57 & 1923.85 & 6.15 & 38.57 & 55.28 \\
Fair-PPO $\alpha=0.5, \beta=1.0$ & 126.09 & 9.96 & 3.99 & 3.61 & 1922.99 & 7.30 & 39.19 & 53.51 \\
Fair-PPO $\alpha=0.75, \beta=0.0$ & 105.59 & 6.88 & 3.94 & 3.55 & 1924.57 & 5.99 & 37.73 & 56.28 \\
Fair-PPO $\alpha=0.75, \beta=0.25$ & 109.51 & 7.24 & 3.97 & 3.60 & 1921.14 & 6.27 & 40.69 & 53.04 \\
Fair-PPO $\alpha=0.75, \beta=0.5$ & 108.14 & 7.64 & 3.95 & 3.56 & 1924.64 & 6.14 & 38.37 & 55.49 \\
Fair-PPO $\alpha=0.75, \beta=0.75$ & 73.94 & 6.77 & 4.07 & 3.77 & 1925.02 & 4.15 & 42.23 & 53.63 \\
Fair-PPO $\alpha=0.75, \beta=1.0$ & 113.40 & 7.87 & 3.96 & 3.59 & 1924.35 & 6.48 & 39.72 & 53.80 \\
Fair-PPO $\alpha=1.0, \beta=0.0$ & 106.34 & 7.40 & 3.95 & 3.56 & 1924.81 & 6.03 & 38.84 & 55.13 \\
Fair-PPO $\alpha=1.0, \beta=0.25$ & 104.19 & 6.89 & 3.94 & 3.56 & 1925.00 & 5.91 & 39.03 & 55.07 \\
Fair-PPO $\alpha=1.0, \beta=0.5$ & 111.74 & 8.45 & 3.95 & 3.57 & 1923.65 & 6.36 & 38.54 & 55.10 \\
Fair-PPO $\alpha=1.0, \beta=0.75$ & 97.13 & 8.25 & 4.05 & 3.73 & 1923.37 & 5.53 & 45.66 & 48.82 \\
Fair-PPO $\alpha=1.0, \beta=1.0$ & 105.70 & 6.80 & 3.95 & 3.57 & 1924.67 & 6.00 & 39.26 & 54.75 \\
\midrule
PPO & 105.92 & 6.83 & 3.94 & 3.56 & 1923.99 & 6.01 & 38.17 & 55.82 \\
\midrule
FEN & 223.76 & 11.74 & 4.54 & 4.09 & 1924.68 & 19.74 & 18.40 & 61.86 \\
\midrule
SOTO $\alpha\text{-fair}=0.9$ & 237.35 & 9.37 & 4.56 & 4.11 & 1312.85 & 26.27 & 17.14 & 56.60 \\
SOTO $\alpha\text{-fair}=1.0$ & 239.60 & 9.21 & 4.51 & 4.06 & 1170.86 & 27.26 & 20.91 & 51.82 \\
SOTO $\alpha\text{-fair}=2.0$ & 239.95 & 9.13 & 4.51 & 4.06 & 1139.43 & 27.67 & 18.47 & 53.86 \\
SOTO $\alpha\text{-fair}=5.0$ & 235.22 & 9.45 & 4.55 & 4.10 & 1172.84 & 25.36 & 17.79 & 56.86 \\
\bottomrule
\end{tabular}}%
\caption{HospitalSim. Average efficiency (patients treated) and fairness performance (demographic disparity) across Fair-PPO and benchmark models. Further metrics are reported to complete the analysis.}
\label{HS-test-results-full}
\end{table*}

\begin{table*}
\centering
\resizebox{\textwidth}{!}{%
\begin{tabular}{l ccc ccc ccc}
\toprule
& \multicolumn{3}{c}{\textbf{Reward Gap}} & \multicolumn{6}{c}{\textbf{Patients Finished}} \\
\cmidrule(lr){2-4} \cmidrule(lr){5-10}
\textbf{Algorithm} & \textbf{High} & \textbf{Medium} & \textbf{Low} & \multicolumn{2}{c}{\textbf{High Prio}} & \multicolumn{2}{c}{\textbf{Medium Prio}} & \multicolumn{2}{c}{\textbf{Low Prio}} \\
\cmidrule(lr){5-6} \cmidrule(lr){7-8} \cmidrule(lr){9-10}
& & & & \textbf{Sens.} & \textbf{Non-Sens.} & \textbf{Sens.} & \textbf{Non-Sens.} & \textbf{Sens.} & \textbf{Non-Sens.} \\
\midrule
Fair-PPO $\alpha=0.0, \beta=0.25$ & 5.57 & 5.59 & 8.48 & 11.37 & 17.63 & 9.13 & 13.85 & 4.49 & 7.17 \\
Fair-PPO $\alpha=0.0, \beta=0.5$ & 5.62 & 5.52 & 8.55 & 11.37 & 17.58 & 9.28 & 14.00 & 4.50 & 7.12 \\
Fair-PPO $\alpha=0.0, \beta=0.75$ & 5.35 & 5.38 & 8.79 & 11.40 & 17.65 & 9.13 & 14.02 & 4.53 & 7.17 \\
Fair-PPO $\alpha=0.0, \beta=1.0$ & 5.24 & 6.02 & 28.84 & 6.06 & 9.59 & 4.61 & 6.91 & 2.33 & 3.58 \\
Fair-PPO $\alpha=0.25, \beta=0.0$ & 5.46 & 5.54 & 8.20 & 11.42 & 17.67 & 9.16 & 14.01 & 4.53 & 7.21 \\
Fair-PPO $\alpha=0.25, \beta=0.25$ & 5.40 & 5.39 & 7.74 & 11.46 & 17.67 & 9.22 & 13.89 & 4.52 & 7.17 \\
Fair-PPO $\alpha=0.25, \beta=0.5$ & 5.66 & 5.53 & 7.49 & 11.40 & 17.64 & 9.18 & 13.82 & 4.51 & 7.04 \\
Fair-PPO $\alpha=0.25, \beta=0.75$ & 5.37 & 5.38 & 8.81 & 11.45 & 17.66 & 9.18 & 13.87 & 4.50 & 7.16 \\
Fair-PPO $\alpha=0.25, \beta=1.0$ & 5.43 & 5.43 & 7.54 & 11.41 & 17.68 & 9.29 & 14.00 & 4.51 & 7.16 \\
Fair-PPO $\alpha=0.5, \beta=0.0$ & 5.43 & 5.39 & 8.39 & 11.47 & 17.67 & 9.24 & 13.84 & 4.48 & 7.12 \\
Fair-PPO $\alpha=0.5, \beta=0.25$ & 5.29 & 5.38 & 8.11 & 11.42 & 17.67 & 9.25 & 14.15 & 4.55 & 7.23 \\
Fair-PPO $\alpha=0.5, \beta=0.5$ & 5.49 & 5.43 & 8.52 & 11.38 & 17.59 & 9.19 & 14.08 & 4.46 & 7.16 \\
Fair-PPO $\alpha=0.5, \beta=0.75$ & 5.50 & 5.45 & 9.27 & 11.38 & 17.66 & 9.23 & 14.00 & 4.51 & 7.16 \\
Fair-PPO $\alpha=0.5, \beta=1.0$ & 5.27 & 5.39 & 9.66 & 11.44 & 17.68 & 9.31 & 14.09 & 4.48 & 7.24 \\
Fair-PPO $\alpha=0.75, \beta=0.0$ & 5.46 & 5.50 & 8.09 & 11.41 & 17.62 & 9.20 & 13.95 & 4.48 & 7.16 \\
Fair-PPO $\alpha=0.75, \beta=0.25$ & 5.26 & 5.36 & 7.92 & 11.46 & 17.68 & 9.25 & 14.04 & 4.51 & 7.25 \\
Fair-PPO $\alpha=0.75, \beta=0.5$ & 5.59 & 5.65 & 8.99 & 11.34 & 17.60 & 9.10 & 13.84 & 4.41 & 7.02 \\
Fair-PPO $\alpha=0.75, \beta=0.75$ & 5.41 & 5.52 & 8.99 & 11.38 & 17.65 & 9.26 & 13.91 & 4.51 & 7.15 \\
Fair-PPO $\alpha=0.75, \beta=1.0$ & 5.45 & 5.47 & 8.39 & 11.46 & 17.69 & 9.17 & 13.87 & 4.53 & 7.14 \\
Fair-PPO $\alpha=1.0, \beta=0.0$ & 5.19 & 5.39 & 7.44 & 11.45 & 17.67 & 9.26 & 14.01 & 4.54 & 7.22 \\
Fair-PPO $\alpha=1.0, \beta=0.25$ & 5.48 & 5.60 & 8.23 & 11.38 & 17.63 & 9.16 & 14.00 & 4.48 & 7.09 \\
Fair-PPO $\alpha=1.0, \beta=0.5$ & 5.43 & 5.45 & 7.63 & 11.47 & 17.60 & 9.19 & 13.95 & 4.45 & 7.09 \\
Fair-PPO $\alpha=1.0, \beta=0.75$ & 5.42 & 5.20 & 8.16 & 11.27 & 18.09 & 9.71 & 15.39 & 5.20 & 8.01 \\
Fair-PPO $\alpha=1.0, \beta=1.0$ & 5.34 & 4.97 & 8.47 & 11.45 & 17.64 & 9.27 & 13.91 & 4.46 & 7.11 \\
\midrule
PPO & 5.32 & 5.30 & 8.10 & 11.47 & 17.67 & 9.22 & 13.89 & 4.48 & 7.12 \\
\midrule
FEN & 11.77 & 11.80 & 11.63 & 27.30 & 47.29 & 27.39 & 47.03 & 27.26 & 47.49 \\
\midrule
SOTO $\alpha\text{-fair}=0.9$ & 9.38 & 9.39 & 9.32 & 29.75 & 49.26 & 30.14 & 49.02 & 29.84 & 49.34 \\
SOTO $\alpha\text{-fair}=1.0$ & 9.28 & 9.18 & 9.19 & 29.96 & 49.57 & 30.55 & 49.57 & 30.17 & 49.78 \\
SOTO $\alpha\text{-fair}=2.0$ & 9.17 & 9.14 & 9.11 & 30.05 & 49.53 & 30.54 & 49.77 & 30.22 & 49.84 \\
SOTO $\alpha\text{-fair}=5.0$ & 9.48 & 9.36 & 9.52 & 29.28 & 48.76 & 29.81 & 48.94 & 29.45 & 48.98 \\
\bottomrule
\end{tabular}}%
\caption{HospitalSim with CSP. Comparison of mean reward gaps and patients finished across different priority levels for Fair-PPO and benchmark models.}
\label{HS-test-results-full-csp}
\end{table*}
\end{document}